\documentclass[aps,pra,twocolumn]{revtex4-1}

\usepackage{graphicx}
\usepackage[english]{babel}
\usepackage{amsmath, amssymb, amsfonts}
\usepackage{bm}
\usepackage[breaklinks]{hyperref}
\hypersetup{
    unicode=true,
    colorlinks=true,
    linkcolor=blue,
    citecolor=blue,
    urlcolor=blue
  }
\usepackage{color}
\definecolor{darkgreen}{rgb}{0.0, 0.3, 0}

\newlength{\figwidth}
\newlength{\lfig}
\newlength{\sfig}
\begin{document}
\title{Near-threshold bound states of the dipole-dipole interaction}
\date{\today}
\author{Tijs Karman}
\affiliation{Joint Quantum Centre (JQC) Durham-Newcastle, Department of
Chemistry, Durham University, South Road, Durham, DH1 3LE, United Kingdom}
\author{Matthew D. Frye}
\affiliation{Joint Quantum Centre (JQC) Durham-Newcastle, Department of
Chemistry, Durham University, South Road, Durham, DH1 3LE, United Kingdom}
\author{John D. Reddel}
\affiliation{Joint Quantum Centre (JQC) Durham-Newcastle, Department of
Chemistry, Durham University, South Road, Durham, DH1 3LE, United Kingdom}
\author{Jeremy M. Hutson}
\affiliation{Joint Quantum Centre (JQC) Durham-Newcastle, Department of
Chemistry, Durham University, South Road, Durham, DH1 3LE, United Kingdom}

\begin{abstract}
We study the two-body bound states of a model Hamiltonian that describes the
interaction between two field-oriented dipole moments. This model has been used
extensively in many-body physics of ultracold polar molecules and magnetic
atoms, but its few-body physics has been explored less fully. With a hard-wall
short-range boundary condition, the dipole-dipole bound states are universal
and exhibit a complicated pattern of avoided crossings between states of
different character. For more realistic Lennard-Jones short-range interactions,
we consider parameters representative of magnetic atoms and polar molecules.
For magnetic atoms, the bound states are dominated by the Lennard-Jones
potential, and the perturbative dipole-dipole interaction is suppressed by the
special structure of van der Waals bound states. For polar molecules, we find a
dense manifold of dipole-dipole bound states with many avoided crossings as a
function of induced dipole or applied field, similar to those for hard-wall
boundary conditions. This universal pattern of states may be observable
spectroscopically for pairs of ultracold polar molecules.
\end{abstract}

\maketitle


\section{Introduction \label{sec:intro}}

The dipole-dipole interaction has been studied extensively in the many-body
physics of ultracold polar molecules and magnetic atoms
\cite{micheli:06,baranov:08,lahaye:09,baranov:12}. It is a long-range,
anisotropic interaction that leads to a range of unique and exotic phenomena:
new quantum phases of matter
\cite{astrakharchik:07,buchler:07,cooper:09,capogrosso-sansone:10,macia:12},
quantum magnetism \cite{barnett:06,gorshkov:11,yan:13,zou:17}, and the
anisotropic collapse of dipolar Bose-Einstein condensates \cite{lahaye:08}.
Dipolar interactions also have applications in quantum computing
\cite{demille:02,lukin:01,yelin:06}, and in quantum simulation
\cite{santos:04,buchler:05,jaksch:05}.

Numerous species with significant dipole moments have now been created at
ultracold temperatures. These include both high-spin magnetic atoms
\cite{Griesmaier:2005, beaufils:2008, Lu:2011, Pasquiou:2012, Aikawa:2012}, and
polar molecules \cite{Ni:KRb:2008, Takekoshi:RbCs:2014, Molony:RbCs:2014,
Park:NaK:2015, Guo:NaRb:2016, Truppe:MOT:2017, Rvachov:2017, McCarron:2018}.
Before exploring the diverse many-body physics experimentally, we need to
understand the 2-body physics, both ultracold scattering above threshold and
weakly bound states below threshold.

The simplest model of the dipole-dipole interaction is that between two dipoles
constrained to be parallel to one another. This model has been used extensively
in many-body physics, but its two-body properties have been studied less fully.
Bohn, Cavagnero and Ticknor (BCT) \cite{ticknor:08,bohn:09} have used this
model for the scattering of two dipoles with orientations fixed in space. They
investigated the low-energy and high-energy regimes, using the Born and eikonal
approximations, respectively. The Born approximation leads to an
energy-independent cross section that scales with the fourth power of the
induced dipole moment. This accounts for all partial waves except for pure
$s$-wave scattering, which requires an additional term involving the scattering
length. Since the scattering length diverges at resonances that occur when
bound states cross threshold, it is important to characterize the bound states
supported by the dipole-dipole interaction. These states may also be accessible
spectroscopically.

In this paper, we investigate the bound states of a simple dipolar Hamiltonian.
Section \ref{sec:hamiltonian} describes the Hamiltonian and the approach used
to compute bound states in the remainder of the paper. Section \ref{sec:adia}
describes the adiabatic potential curves produced by this Hamiltonian, and
estimates the number of bound states supported by these curves semiclassically.
Section \ref{sec:hardwall} discusses the bound states supported by the
dipole-dipole Hamiltonian with hard-wall short-range boundary conditions.
Section \ref{sec:LJ} employs a more realistic Lennard-Jones short-range
interaction and discusses the bound states for two sets of parameters,
representative of magnetic atoms and polar molecules, respectively.

\section{Hamiltonian \label{sec:hamiltonian}}

The Schr{\"o}dinger equation considered by BCT \cite{bohn:09} is
\begin{align}
\left[ -\frac{\hbar^2}{2M} \nabla^2 + \hat{V}_\mathrm{dip}(\bm{R})
+ V_\mathrm{SR}(R) \right] \psi = E \psi. \label{eq:bct}
\end{align}
Here, $M$ is the reduced mass,
$V_\mathrm{SR}(R)$ is a short-range interaction,
and
\begin{align}
\hat{V}_\mathrm{dip}(\bm{R}) = -\frac{2d_1d_2}{4\pi \epsilon_0} \frac{P_2(\cos\theta)}{R^3}
\label{eq:fixdip}
\end{align}
is the anisotropic dipole-dipole interaction. This Hamiltonian describes the
interaction of two electric dipole moments, of magnitude $d_1$ and $d_2$, with
fixed orientations in space. This commonly arises when both dipoles are
oriented along the same space-fixed field. The distance between the dipole
moments is $R$ and the angle between the interdipole axis and the external
field axis is $\theta$. The interaction between two magnetic dipoles $\mu$ also
follows Eq.\ \ref{eq:fixdip} if expressed in terms of effective electric
dipoles $d=\mu/c$, where $c$ is the speed of light.

Conceptually, the long-range dipole-dipole and short-range interactions act on
different length scales. The short-range interaction may be
viewed as defining a boundary condition
for the pure dipole-dipole problem, obtained by dropping $V_\mathrm{SR}(R)$.
The dipole-dipole part of the Schr\"odinger equation can then be
brought into dimensionless form \cite{ticknor:08,bohn:09}
\begin{align}
\left[ -\frac{1}{2} \frac{d^2}{dr^2} + \frac{\hat{L}^2}{2 r^2} -
\frac{2P_2(\cos\theta)}{r^3} \right] \psi = \varepsilon \psi,
\label{eq:seuniv}
\end{align}
where $r=R/R_\mathrm{dip}$, $\varepsilon=E/E_\mathrm{dip}$, and the dipole
length and dipole energy are defined
\cite{bohn:09} as
\begin{align}
R_\mathrm{dip} &= \frac{M d_1d_2}{\hbar^2 4 \pi \epsilon_0}, \nonumber \\
E_\mathrm{dip} &= \frac{\hbar^2}{MR_\mathrm{dip}^2}
= \frac{\hbar^6 (4\pi\epsilon_0)^2}{M^3 d_1^2d_2^2}.
\label{eq:scales}
\end{align}
It should be stressed that the dipole energy defined by Eq.\ \ref{eq:scales} is
\emph{not} a measure of the dipole-dipole interaction energy, as the name may
suggest. In fact, the dipole energy decreases as the dipole moment increases,
whereas the strength of the dipole-dipole interaction increases. Because of
this, the set of dipole-dipole bound states becomes more dense as the dipole
moment increases.

The dimensionless Schr{\"o}dinger equation, Eq.~\eqref{eq:seuniv}, leads to
dynamics that is universal in the sense that it may be scaled for the
parameters of any particular system \cite{bohn:09} and depends only on a
short-range boundary condition. In ultracold scattering, many systems follow a
stronger form of universality, where scaled properties depend only on the
$s$-wave scattering length \cite{Braaten:2006}, with no further dependence on
different boundary conditions that generate the same scattering length.
However, the dipole-dipole interaction is anisotropic and couples different
partial waves. The dynamics thus depends on multiple partial waves, which may
not exhibit the same periodicity with s-wave scattering length. Dipole-dipole
systems thus do not follow the stronger form of universality.

Here, we model the short-range interaction or boundary condition in two
different ways. First, we consider a hard-wall boundary condition and explore
the near-threshold bound states as a function of the hard-wall position. This
constitutes the simplest model of dipole-dipole interactions. Secondly, we
include a Lennard-Jones short-range potential, which can model van der Waals
attraction and short-range repulsion more physically. In particular, this
introduces correlations between the short-range phases for different partial
waves more physically than placing a hard wall at identical separations for all
partial waves.

\subsection{Computational Methods \label{sec:method}}

We perform coupled-channels calculations to obtain bound states, using the {\sc
bound} computer code \cite{Hutson:bound:2011}. The nuclear wave function is
expanded in partial waves, represented by the spherical harmonics
$Y_{LM_L}(\theta,\phi)$. These are coupled by the dipole-dipole interaction,
with matrix elements \cite{brink}
\begin{align}
& \left\langle L M_L \left| \frac{-2P_2(\cos\theta)}{r^3}
\right| L' M'_L \right\rangle = -2 r^{-3} (-1)^{M_L'}\\ \nonumber
& \times \sqrt{(2L+1)(2L'+1)}
\begin{pmatrix}
L & 2 & L' \\
-M_L & 0 & M_L'
\end{pmatrix}
\begin{pmatrix}
L & 2 & L' \\
0 & 0 & 0
\end{pmatrix},
\end{align}
which is non-zero only if $M_L=M_L'$ and $|L-L'|= 2$ or 0 (but not if
$L=L'=0$). We mainly consider the case $M_L=0$ for even $L$, for which the
$s$-wave $L=0$ channel occurs.

The solution of the coupled equations is propagated in two steps. The diabatic
modified log-derivative propagator of Manolopoulos \cite{manolopoulos:86} is
used outwards on an equidistant grid of spacing $\Delta R = 10^{-5}
R_\mathrm{dip}$ from $R_\mathrm{min}$ to $R_\mathrm{mid}=0.2 R_\mathrm{dip}$.
The Airy propagator of Alexander and Manolopoulos \cite{alexander:87} is used
to propagate inwards from $R_\mathrm{max}=3R_\mathrm{dip}$ to $R_\mathrm{mid}$
using a radial grid with variable step size. The boundary conditions are such
that the wave function vanishes at $R_\mathrm{min}$ and follows a
Wentzel-Kramers-Brillouin (WKB) form at $R_\mathrm{max}$. At the matching
point, $R_\mathrm{match}$, bound states are found by locating a zero in an
eigenvalue of the matching matrix, \emph{i.e.}, the difference of incoming and
outgoing log-derivative matrices, as a function of energy
\cite{Hutson:CPC:1994}. The matching point does not coincide with
$R_\mathrm{mid}$, but is chosen at shorter separation to ensure that the
matching is performed in the classically allowed region. In the case of
hard-wall boundary conditions, the matching point is chosen as
$R_\mathrm{match}=R_\mathrm{min}+10^{-3} R_\mathrm{dip}$, and otherwise we use
$R_\mathrm{match}=8~a_0$, which is close to the minimum of the Lennard-Jones
potential.

\section{Adiabatic potential curves\label{sec:adia}}

First, we consider the adiabatic potential curves (adiabats) of the reduced
Hamiltonian in Eq.~\ref{eq:seuniv}. The adiabats $\varepsilon_n^{\rm ad}(r)$
are defined as the eigenvalues of
\begin{align}
\frac{\hat{L}^2}{2r^2} - \frac{2P_2(\cos\theta)}{r^3}
\end{align}
for fixed $r = R/R_\mathrm{dip}$. For $r \gg 1$, the centrifugal term dominates
the dipole-dipole interaction, and the adiabatic states correspond to spherical
harmonics $Y_{LM_L}(\theta,\phi)$. The lowest adiabat corresponds
asymptotically to $L=0$ and has a vanishing first-order dipole-dipole
interaction. Dipole-dipole coupling to the $L=2$ channel yields the
second-order energy
\begin{align}
\varepsilon^{(2)}_{L=0}(r) & = -\frac{r^2}{3} \left| \left\langle 0 0
\left| \frac{2P_2(\cos\theta)}{r^3} \right| 2 0 \right\rangle \right|^2 \nonumber \\
& = -\frac{4}{15} r^{-4}.
\end{align}
The lowest adiabat thus varies asymptotically as $-C_4R^{-4}$. We can define a
corresponding length scale as
\begin{align}
R_4 = \sqrt{\frac{2M C_4}{\hbar^2}} = \sqrt{\frac{8}{15}} R_\mathrm{dip} \approx 0.73 R_\mathrm{dip}.
\end{align}
This differs slightly from the value 1.09~$R_\mathrm{dip}$ given in
Ref.~\onlinecite{bohn:09}.
In the opposite limit, where the dipole-dipole
interaction dominates, the lowest adiabat is simply $-2 r^{-3}$ and the
corresponding eigenstate is localized completely at $\theta=0$ and $\pi$, where
the dipoles lie head-to-tail along the field direction.

\begin{figure}
\begin{center}
\includegraphics[width=0.9\lfig,clip]{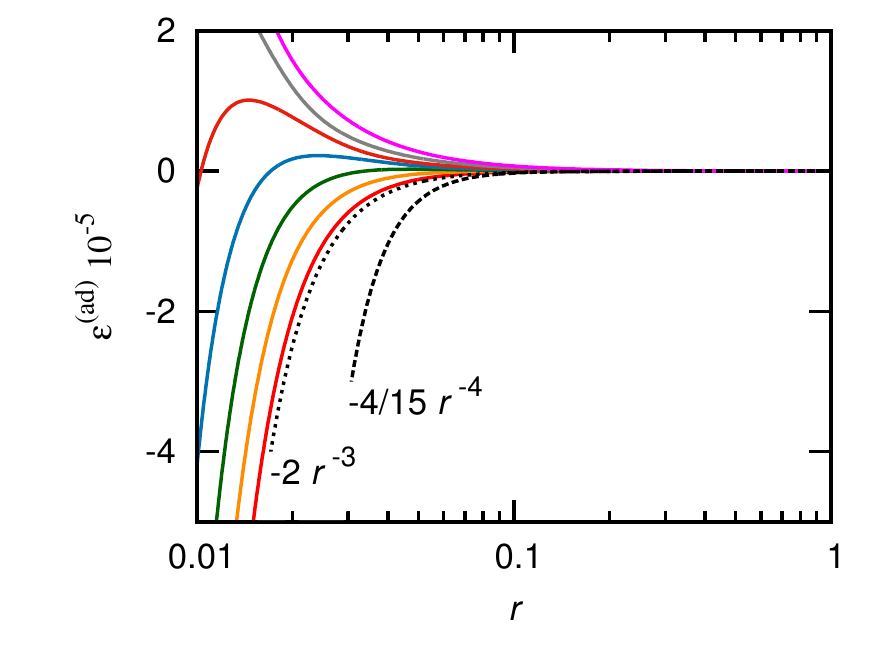}
\includegraphics[width=0.9\lfig,clip]{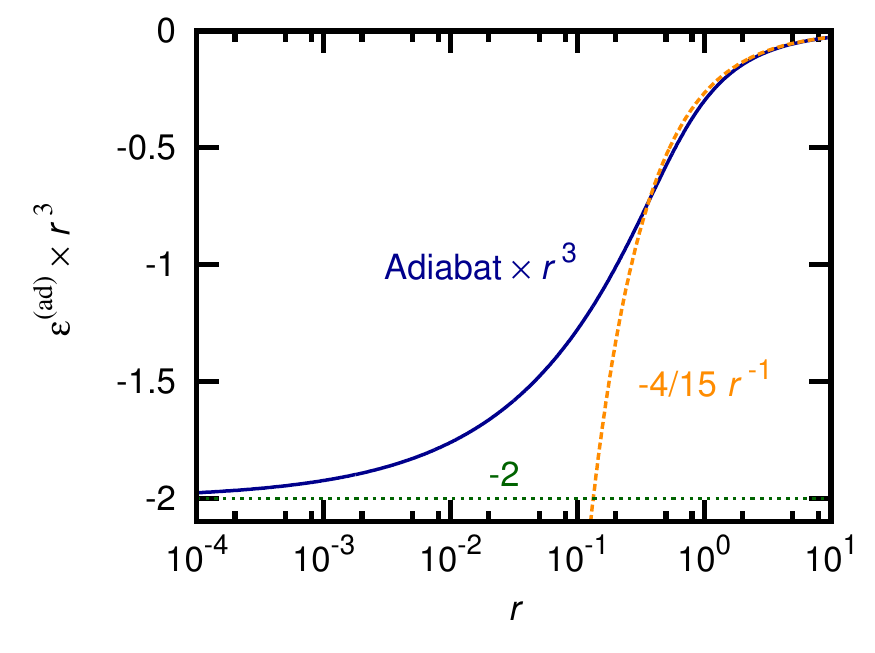}
\includegraphics[width=0.9\lfig,clip]{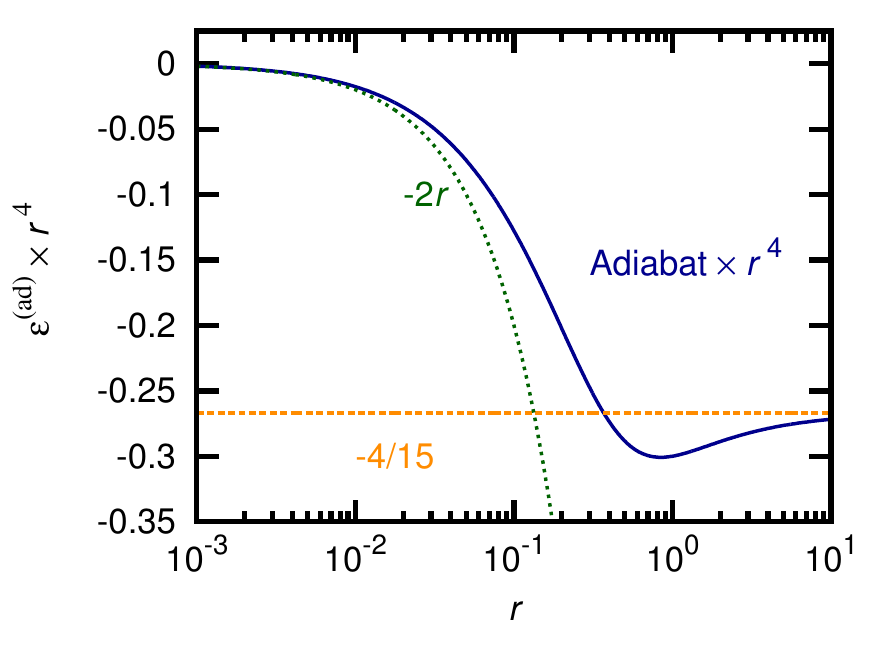}
\caption{ \label{fig:adiabats}
Top: Low-lying adiabats of the BCT Hamiltonian, Eq.~\eqref{eq:seuniv}, for even $L$
and $M_L=0$. Middle and bottom: Lowest adiabat multiplied by $r^3$ and $r^4$ respectively.
Short-range and long-range approximations to the lowest adiabat, $-2r^{-3}$ and $-4/15
r^{-4}$, respectively, are also included.
}
\end{center}
\end{figure}

The calculated adiabats are shown in Fig.~\ref{fig:adiabats}. The lower two
panels of this figure show the lowest adiabat multiplied by $r^3$ and $r^4$,
and show how it approaches the short-range and long-range limits $-2 r^{-3}$
and $-4/15 r^{-4}$. There is clearly an extended region for intermediate $r$
where neither approximation is accurate. The long-range $r^{-4}$ form is
accurate only for $r>10$. Therefore, this form is appropriate for describing
states only if they are bound by a small fraction of the dipole energy; in many
cases there are no such states. This contrasts with other long-range
interactions, such as the $R^{-6}$ interaction between neutral atoms, which
often remain appropriate at depths of hundreds to thousands of times their
corresponding energy scale. When $r$ is of order unity, the dipole-dipole
coupling and centrifugal terms are roughly comparable. This leads to
nonadiabatic coupling as the eigenstates change with $r$ from freely orbiting
to states localized in $\theta$ by the dipole-dipole potential. Describing the
states in a partial-wave expansion requires inclusion of functions with
increasing values of $L$ as $r$ decreases. The change in character takes place
over an extended range of $r$, as the relative strength of the two terms
depends only linearly on $r$.

Higher-$L$ adiabats are repulsive at long range, and have centrifugal barriers
that move inwards and increase in height with increasing $L$. Outside the
barrier, the potentials are dominated by the centrifugal term. Inside the
barrier, the dipole-dipole interaction dominates; eventually the eigenstates
again localize in $\theta$, leading to adiabatic potentials proportional to
$r^{-3}$.

\subsection{WKB estimate of the number of bound states}

\begin{figure}
\begin{center}
\includegraphics[width=\lfig,clip]{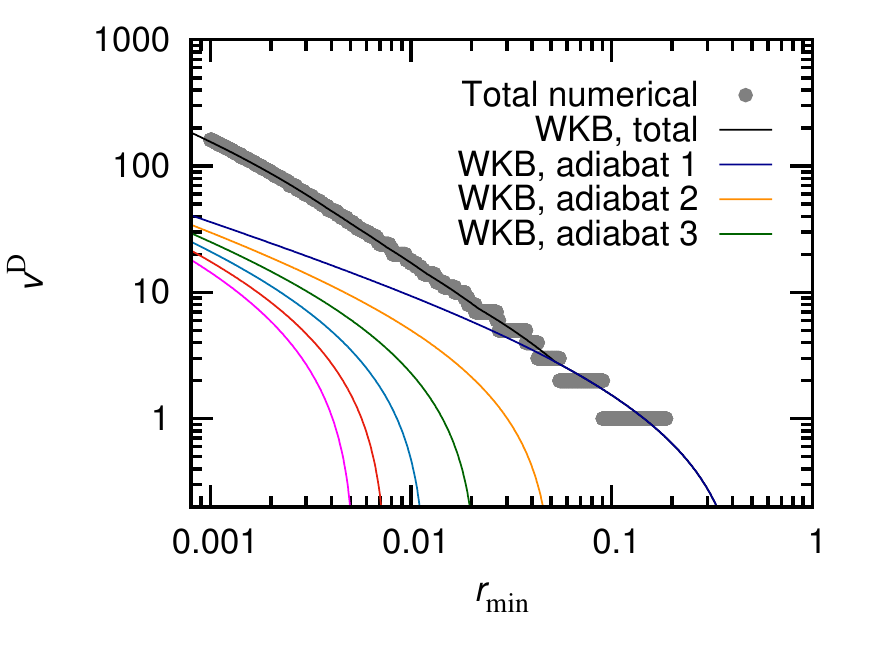}
\caption{ \label{fig:BCT_WKB_BOUND}
The total number of states obtained from coupled-channels calculations compared
to the semi-classical WKB result. }
\end{center}
\end{figure}

In this section, we estimate the number of bound states supported by each
adiabat semiclassically. To this end, we compute the phase integral
at zero energy
\begin{align}
\Phi_n(r_\mathrm{min}) = \int_{r_\mathrm{min}}^\infty \Re
\left(\sqrt{2\varepsilon_n^{\rm ad}(r)}\right) dr,
\end{align}
for each adiabat.
Taking the real part of the
integrand ensures that only the classically allowed region contributes. The WKB
quantization condition is used to define a non-integer quantum number at
dissociation, \emph{i.e.}, at zero energy,
\begin{align}
v^{\rm D}_{n}(r_\mathrm{min}) = \frac{\Phi_{n}(r_\mathrm{min})}{\pi}-\frac{1}{2}
\end{align}
which serves to estimate the total number of bound states.

Estimated numbers of bound states for different adiabats, and the total number
of bound states, are shown in Fig.~\ref{fig:BCT_WKB_BOUND}. The multichannel
node count \cite{Johnson:1978} at zero energy from coupled-channels
calculations is also shown for comparison. As $r_\mathrm{min}$ is decreased,
the number of bound states supported by the lowest adiabat increases rapidly.
Furthermore, excited adiabats start to contribute bound states as
$r_\mathrm{min}$ is decreased to include the negative-energy regions of these
adiabats, inside their long-range centrifugal barriers. The first states appear
with $r_\mathrm{min}$ of order 0.1, and it can be seen from Fig.\
\ref{fig:adiabats} that in this region the lowest adiabat has deviated
substantially from its $r^{-4}$ long-range form. Therefore, for our purposes,
it is clearly insufficient to approximate the energy of the lowest adiabat by
its long-range form. For $r_\mathrm{min} \ll 0.1$, the adiabatic states
localize in $\theta$ as the $r^{-3}$ dipole-dipole interaction dominates, and
the number of states supported by each adiabat approaches the corresponding
power-law dependence $v^{\rm D}_{n} \propto r_\mathrm{min}^{-1/2}$. The total
number of bound states rises as a higher inverse power than the number in the
individual adiabats as the number of contributing adiabats also rises rapidly
as $r_\mathrm{min}$ decreases.

\section{Hard-wall boundary condition \label{sec:hardwall}}

\begin{figure}
\begin{center}
\includegraphics[width=\lfig,clip]{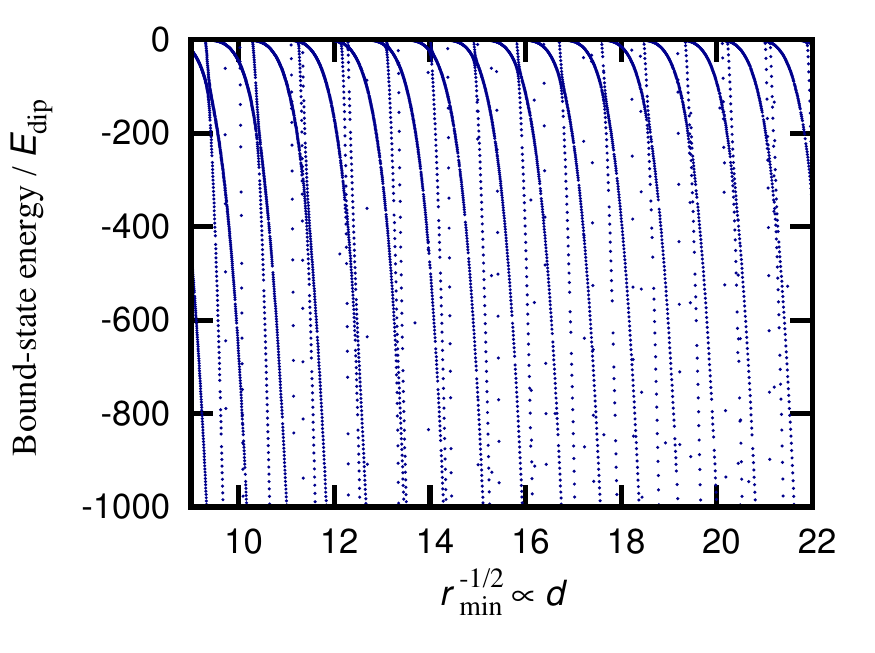}
\includegraphics[width=\lfig,clip]{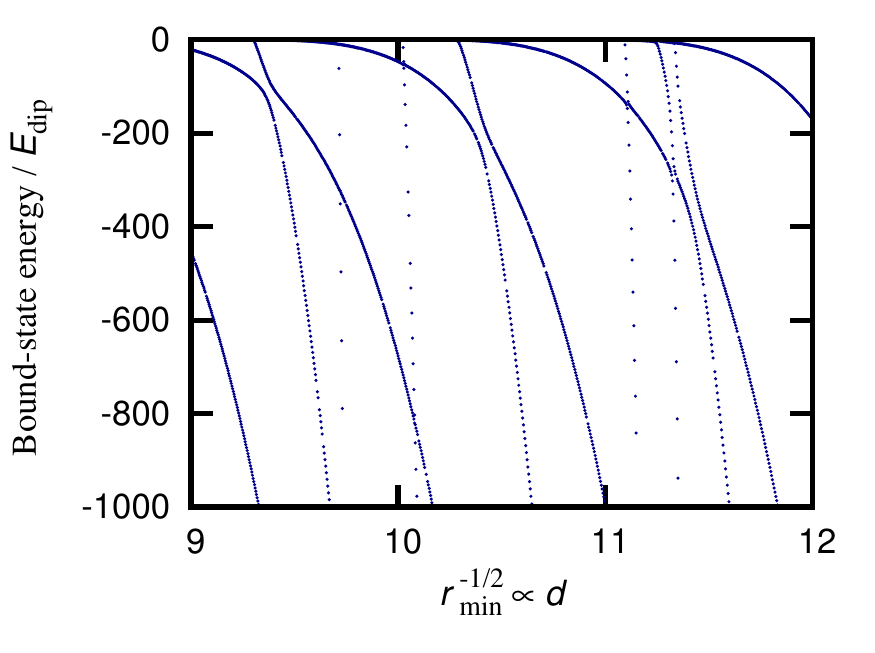}
\caption{ \label{fig:BCT_bound_dip}
Bound states as a function of $r_\mathrm{min}^{-1/2}$, which is
proportional to the dipole moment for fixed short-range boundary condition.}
\end{center}
\end{figure}

We calculate dipole-dipole bound states as a function of the position of the
hard-wall short-range boundary condition, $R_\mathrm{min}$. These calculations
use coupled-channel calculations as described in section \ref{sec:method}, and
do not make an adiabatic separation. The bound-state energies are universal
functions of $r_\mathrm{min}=R_\mathrm{min}/R_\mathrm{dip}$. They are shown in
Fig.\ \ref{fig:BCT_bound_dip} as a function of $r_\mathrm{min}^{-1/2}$, which
is proportional to the dipole moment for fixed $R_\mathrm{min}$. This figure
can therefore be viewed as showing bound states as a function of the dipole
moment. It corresponds to the way that polar molecules could be controlled by
varying an applied electric field: The short-range boundary condition is fixed
by the short-range potential, while the induced dipole moment varies with
field. There is a regular series of states tending relatively slowly towards
threshold, which will create broad resonances in the s-wave dipolar scattering
\cite{Ticknor:long-range:2005}. These are crossed by steeper states which will
create additional narrower resonances.

\begin{figure}
\begin{center}
\includegraphics[width=\lfig,clip]{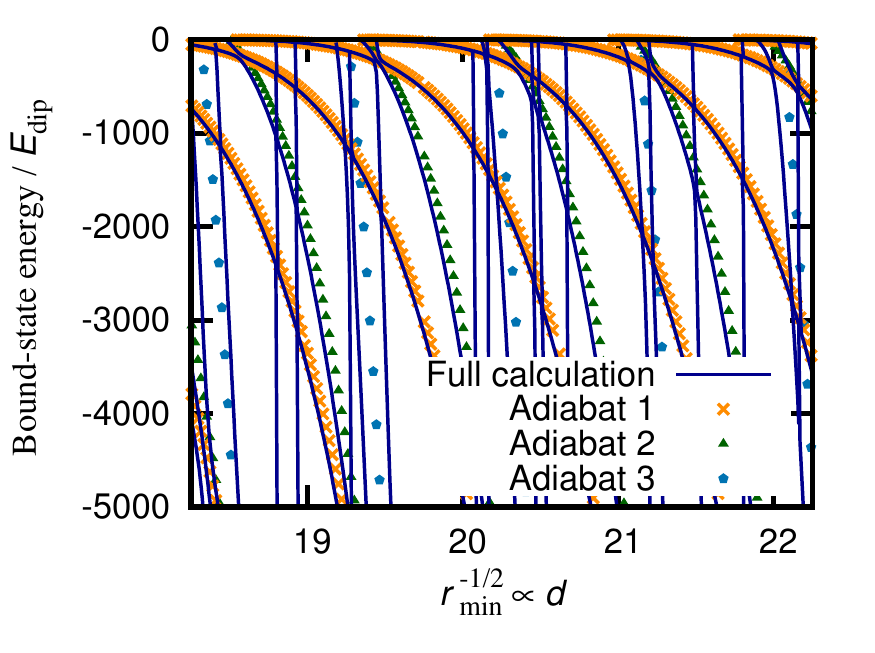}
\caption{ \label{fig:BCT_bound_adiabatic}
Bound states as a function of $r_\mathrm{min}^{-1/2}$,
both from adiabatic and full coupled-channels calculations.
}
\end{center}
\end{figure}

Bound states calculated in the adiabatic approximation are compared to the full
calculation in Fig.~\ref{fig:BCT_bound_adiabatic}. The adiabatic calculations
agree well with the coupled channels calculations except near avoided
crossings, demonstrating that the dynamics is mostly adiabatic. The steeper
states are supported by excited adiabats, which have increasingly high barriers
at large interdipole distances that confine the wavefunction to relatively
short range and increase the vibrational spacing at threshold. This explains
the difference in dependence on and periodicity with $r_\mathrm{min}$ for the
different types of states observed. This difference in periodicity can be
clearly seen in the energy of the avoided crossing between the states in the
lowest two adiabats, which shifts slightly between different repetitions of the
pattern. Because of this, the states are not completely determined by the
s-wave scattering length. This may be viewed as a breakdown of the stronger
form of universality described in section \ref{sec:hamiltonian}. The dipole
moment can be tuned to anywhere in the periodic structure of resonances by
modest changes in the induced dipole moment, provided that the system has a
large dipole length $R_\mathrm{dip} \gg R_\mathrm{min}$. If for example
$R_\mathrm{dip}/R_\mathrm{min} = 100$, which is achievable for polar molecules
\cite{bohn:09}, tuning through a period of the scattering length for the lowest
adiabat requires varying the dipole moment by $10~\%$.

\begin{figure}
\begin{center}
\includegraphics[width=\lfig,clip]{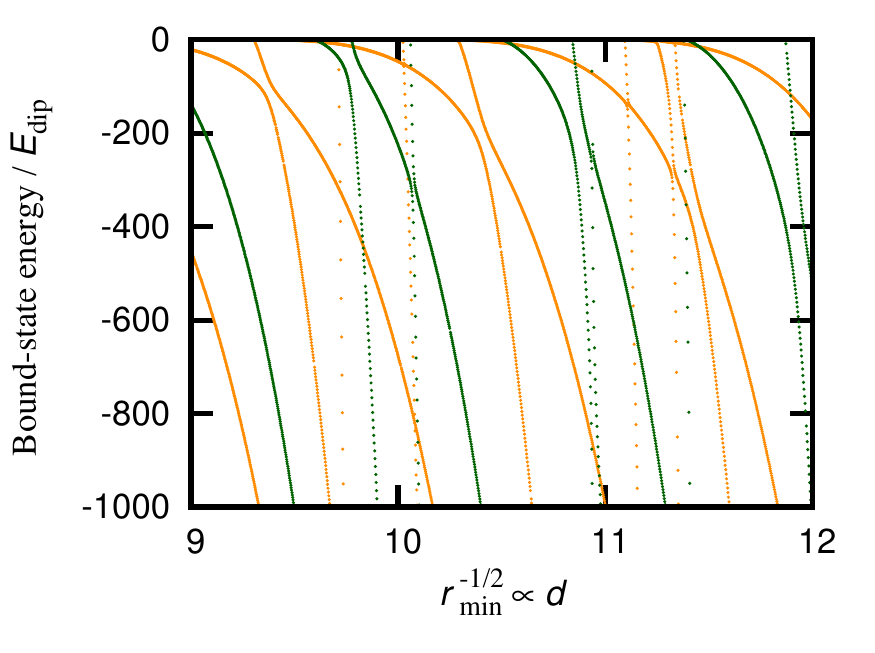}
\caption{ \label{fig:BCT_bound_dip_L}
Bound states as a function of $r_\mathrm{min}^{-1/2}$, for odd $L$.
Results shown in orange and green correspond to $M_L=0$ and $|M_L|=1$, respectively. }
\end{center}
\end{figure}

\begin{figure}
\begin{center}
\includegraphics[width=\lfig,clip]{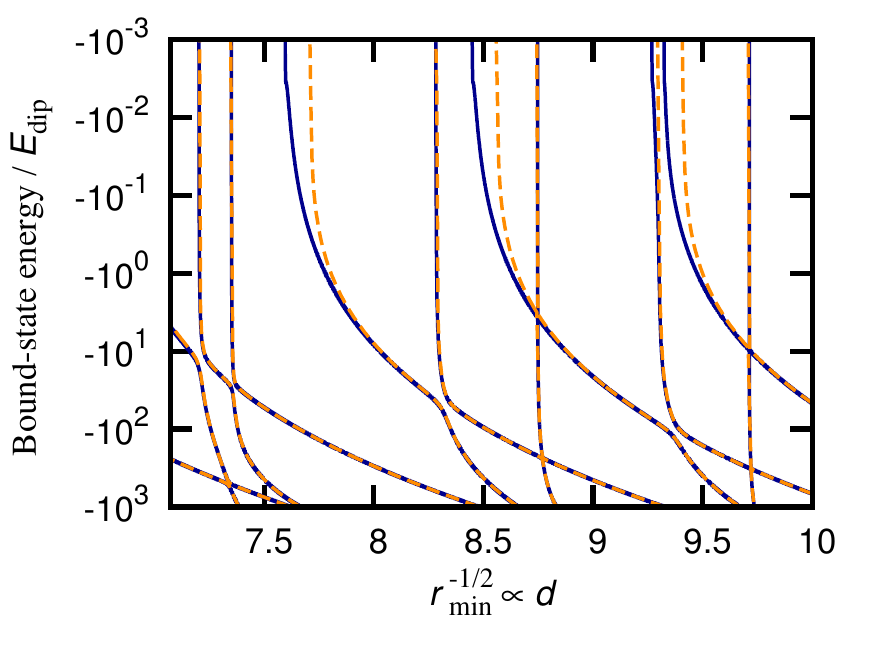}
\caption{ \label{fig:BCT_bound_dip_L2}
Bound states as a function of $r_\mathrm{min}^{-1/2}$, for even and odd $L$ and
$M_L=0$ shown in blue and orange, respectively. }
\end{center}
\end{figure}

We next consider the case of identical fermions. In this case exchange symmetry
requires that $L$ is odd, and states with both $M_L=0$ and $|M_L|=1$ can cause
resonances at the p-wave ($L=1$) threshold. $M_L$ is a conserved quantity, so
states with different values of $M_L$ can cross.
Figure~\ref{fig:BCT_bound_dip_L} shows the resulting bound states as a function
of $r_\mathrm{min}^{-1/2} \propto d$. States shown in orange correspond to
$M_L=0$, whereas those shown in green correspond to $|M_L|=1$. The $M_L=0$
states are very close to those obtained in the bosonic case, shown in Fig.
\ref{fig:BCT_bound_dip}, even though they come from an apparently very
different calculation. The bound states for bosons and fermions with $M_L=0$
are compared in Fig.~\ref{fig:BCT_bound_dip_L2}. The two sets of results are
almost identical for states bound by more than $E_\mathrm{dip}$. Fermion states
for $|M_L|=1$ approach threshold more steeply with $R_\mathrm{min}$ than the
corresponding states for $M_L=0$; this is because they have a larger barrier,
due to a repulsive first-order dipole-dipole interaction.

\begin{figure}
\begin{center}
\includegraphics[width=\lfig,clip]{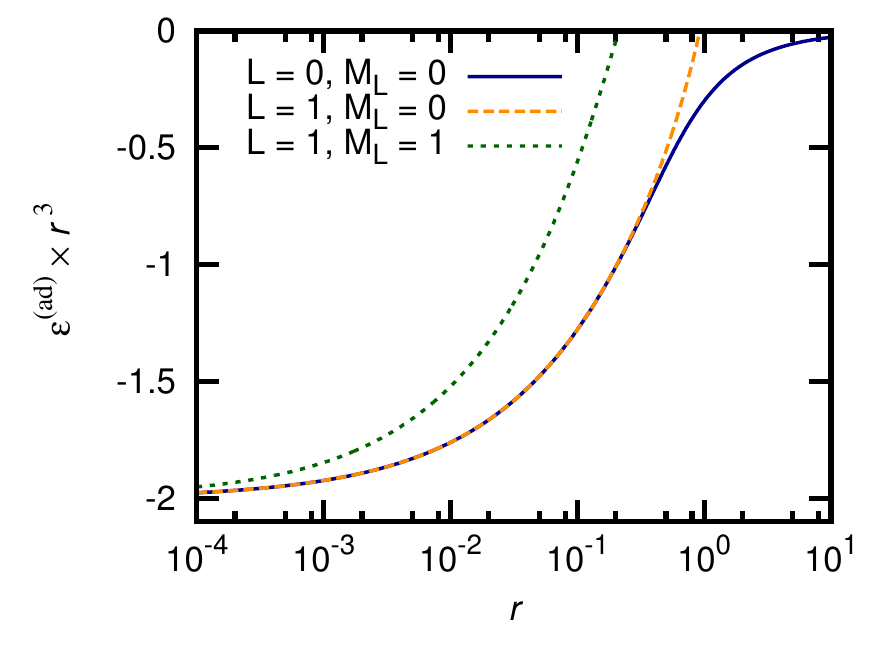}
\caption{ \label{fig:adiabats_fermions}
Adiabatic potential curves multiplied by $r^3$ for bosons ($L=0$) and fermions
($L=1$) for both $M_L=0$ and $1$. }
\end{center}
\end{figure}

We compare the adiabats for bosons and fermions in Fig.\
\ref{fig:adiabats_fermions}. The lowest adiabat for fermions has $M_L=0$ and
shows the same short-range behavior as for bosons, with limit $-2r^{-3}$. Even
though fermion states with $M_L=0$ have nodes at $\theta=\pi/2$, they localize
at $\theta=0$ and $\pi$, in the same way as boson states, when the
dipole-dipole interaction dominates. The lowest adiabats for bosons and
fermions nevertheless differ asymptotically, where the dipole-dipole
interaction is weak enough that the region around $\theta=\pi/2$ is sampled;
these differences become important for $r> 0.3$. This is why differences
between the bound states emerge when they are bound by less than
$E_\mathrm{dip}$.

\section{Lennard-Jones potential \label{sec:LJ}}

In this section we employ a soft-wall boundary condition. We use a
Lennard-Jones ``short-range'' potential in Eq.\ \ref{eq:bct},
\begin{align}
V_\mathrm{SR}(R) = C_{12} R^{-12} - C_6 R^{-6}.
\end{align}
The $R^{-6}$ term is typically attractive and can model the dispersion or van
der Waals interaction, whereas the $R^{-12}$ term describes short-range
repulsion. In contrast with the dipole-dipole interaction, this short-range
potential is completely isotropic. To adjust the short-range behavior, we vary
the Lennard-Jones potential well depth,
\begin{align}
D_{\rm e} = \frac{C_6^2}{4C_{12}},
\end{align}
while holding $C_6$ fixed. A length scale for the van der Waals interaction can
be defined as
\begin{align}
R_6 &= \left( \frac{ 2M C_6 }{\hbar^2} \right)^{1/4},
\end{align}
with energy scale
\begin{equation}
E_6=\frac{\hbar^2}{2MR_6^2}.
\end{equation}
For pairs of magnetic atoms, $R_6$ is comparable to the dipole length,
whereas for pairs of polar molecules the dipole length is considerably larger.

Inclusion of the Lennard-Jones potential breaks the universality of the
dipole-dipole interaction, as the bound states depend on the relative strength
of the dispersion and dipole-dipole interactions. Below, we consider parameters
that are typical for two cases: pairs of magnetic atoms, and pairs of polar
molecules.

\subsection{Magnetic atoms: $R_\mathrm{dip} \approx R_6$}

Here we employ the parameters $C_6 = 2003\, E_{\rm h}a_0^6$
\cite{Maier:ChaosErDy:2015}, a magnetic dipole moment $\mu=9.93\mu_{\rm B}$,
and reduced mass $M = 81.96\,m_{\rm u}$, which correspond to bosonic
$^{162}\mathrm{Dy}$. The Lennard-Jones well depth is varied around $D_{\rm e}
\approx 800$~cm$^{-1}$ \cite{Petrov:2012}, such that it supports around 60
vibrational states for $L=0$. The length scales of the van der Waals, $R_6 =
154~a_0$, and dipole-dipole interaction, $R_\mathrm{dip}=196~a_0$, are roughly
comparable. The Lennard-Jones potential nevertheless supports many more bound
states than the dipole-dipole potential, because it is substantially deeper at
short range.

\begin{figure*}
\begin{center}
\includegraphics[width=0.95\textwidth,clip]{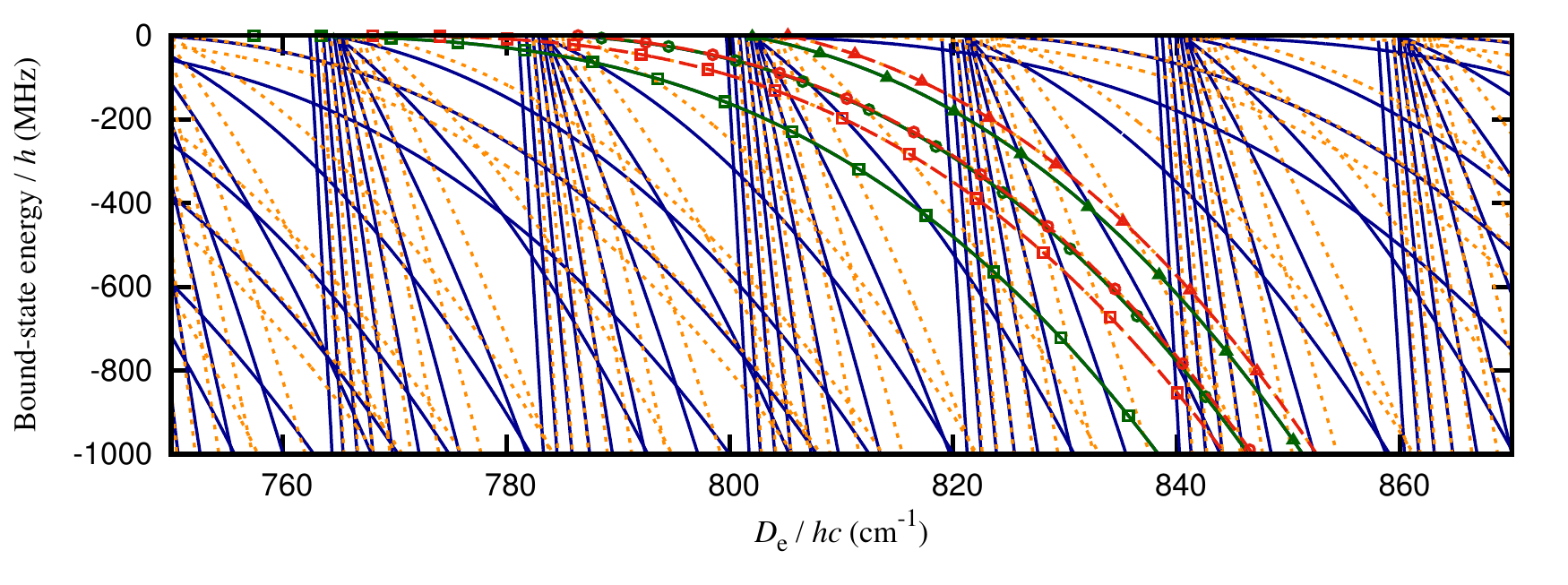}
\caption{ \label{fig:Dy01} Bound states as a function of the Lennard-Jones well
depth. States shown as solid blue lines have been obtained with parameters that
are representative of magnetic atoms. Results shown as orange dashed lines are
obtained with the dipole moment set to zero. Three states from each calculation
are highlighted to guide the eye: in green (red) with the dipole on (off);
squares are states on the lowest adiabat (asymptotically $L=0$), circles are
states on the first excited adiabat ($L=2$), and triangles are states on the
second excited adiabat ($L=4$).}
\end{center}
\end{figure*}

Figure~\ref{fig:Dy01} displays bound states for even $L$ and $M_L=0$ as a
function of the Lennard-Jones well depth, $D_e$. States shown in blue are
from the full calculation, as described above, whereas those shown in orange were
are for the pure Lennard-Jones potential without dipole-dipole interactions.
The structure of the bound states of the full system closely resembles that of
the pure Lennard-Jones potential, and the effect of the dipole-dipole
interaction is essentially perturbative.

\begin{figure}
\begin{center}
\includegraphics[width=1.15\lfig,clip]{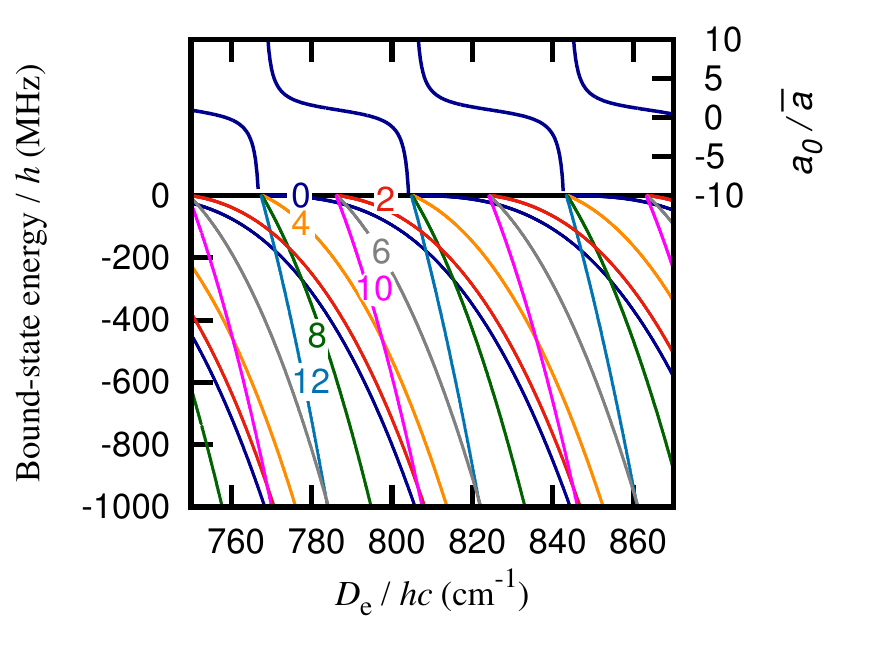}
\caption{ \label{fig:Dy0L}
Bound states, including $L$ assignment, as a function of the Lennard-Jones well
depth in the absence of dipole-dipole interactions. Also included is the
$s$-wave scattering length, which sets the observed period.}
\end{center}
\end{figure}

Figure~\ref{fig:Dy0L} shows an expanded plot of the pure Lennard-Jones states
with $L$ assignments, together with the $s$-wave scattering length. The
structure of the near-threshold bound states of the pure Lennard-Jones
potential is such that it groups together near-threshold states with
$L=0,4,8,\ldots$, and the same holds for states with $L=2,6,10,\ldots$
\cite{Gao:2000, gao:2004:rotational}. The bound states show periodicity as a
function of the well depth, where bound states with $L=0,4,8,\ldots$ cross the
dissociation threshold for well depths where the $s$-wave scattering length $a$
is infinite \cite{Gao:2000, gao:2004:rotational}. Bound states with
$L=2,6,10,\ldots$ cross threshold where $a$ is equal to the mean scattering
length \cite{Gao:2000, gao:2004:rotational}, $a = \bar{a} \approx 0.478 R_6$
\cite{Gribakin:1993}. Away from threshold, the bound states for different $L$
separate, varying more steeply with well depth for higher $L$.

This level structure leads to a grouping of states with $\Delta L \ge 4$,
whereas the dipole-dipole coupling is non-zero only between states with $\Delta
L =0,\pm 2$. This leads to a suppression of the effects of dipole-dipole
coupling; crossings of Lennard-Jones states directly coupled by the
dipole-dipole interaction do not occur near threshold \footnote{In the presence
of an anisotropic dispersion interaction, channels with different values of $L$
may have different effective $C_6$ coefficients; this would lift the threshold
degeneracy between states with $\Delta L=4$ and might result in crossings
between states with $\Delta L=2$ that are coupled by the dipole-dipole
interaction.}. The main effect of dipole-dipole interaction is a shift of the
bound-state energies, which is close to the first-order energy
\begin{align}
E_{vLM_L}^{(1)} =& - 2 d_1d_2
\begin{pmatrix}
L & 2 & L \\
0 & 0 & 0
\end{pmatrix}
\begin{pmatrix}
L & 2 & L \\
M_L & 0 & -M_L
\end{pmatrix}
\nonumber \\
&\times  (-1)^{M_L}(2L+1) \langle vLM_L | R^{-3} | vLM_L\rangle.
\end{align}
The $L=0$ and $L=2$ states are notable exceptions. The $L=0$ states have no
first-order shifts, whereas each $L=2$ state is shifted down in first order to
near-degeneracy with the corresponding $L=0$ state. Higher-order couplings
shift the $L=0$ states down considerably, and shift the $L=2$ states back up,
coincidentally for this particular dipole moment to near the unperturbed
Lennard-Jones level.

\subsection{Molecule-molecule: $R_\mathrm{dip} \gg R_6$}

In this section, we perform calculations where the space-fixed dipole moment is
scaled up by a factor of 10, which results in the equivalent of a space-fixed
electric dipole moment of 0.92 Debye. This increases the dipole length
$R_\mathrm{dip}$ by a factor 100, to 19600 $a_0\approx 1$ $\mu$m, and decreases
the dipole energy $E_\mathrm{dip}$ by a factor of 10000, to 57~kHz~$\times\ h$.
In this case, the van der Waals potential provides a ``short-range'' boundary
condition for the dipole-dipole coupling, which acts on a much larger length
scale. The case $R_\mathrm{dip} \approx 100 R_6$ is roughly typical for
ultracold polar molecules, where the dispersion coefficient is dominated by the
rotational contribution, $C_6 = \mu^4/[(4\pi\epsilon_0)^2\ 6B_\mathrm{rot}]$.
Table~\ref{tab:molecules} gives the molecule-fixed dipole moments
$d^\mathrm{lim}$ and rotational $C_6$ coefficients for selected alkali-metal
dimers. Also given are dispersion and dipole-dipole length scales, with the
latter ($R_\mathrm{dip}^\mathrm{lim}$) calculated with the molecule-fixed
dipole moment, and the electric field $F_\mathrm{0.7}$ at which the space-fixed
dipole moment is approximately 0.7 times its molecule-fixed value and
$R_\mathrm{dip}\approx R_\mathrm{dip}^\mathrm{lim}/2.$

\setlength{\tabcolsep}{8pt}
\begin{table*}
\caption{ Molecular dipole moments and rotational $C_6$ coefficients for
selected alkali dimers. Also included are the dipole and dispersion length
scales, and their ratio. The dipole length is calculated for the limiting value
of the space-fixed dipole moment. The applied electric field $F_\mathrm{0.7}$
needed to induce a dipole moment of $0.7$ times the limiting dipole moment --
and to achieve approximately half the limiting dipolar length scale -- is also
included. \label{tab:molecules}}
\begin{tabular}{lcccccccc}
\hline\hline
Molecule & $d^\mathrm{lim}$ (Debye) & $R_\mathrm{dip}^\mathrm{lim}$ / $a_0$ & $E_\mathrm{dip}^\mathrm{lim}$ (Hz) & $C_6 / E_{\rm h} a_0^6$ & $R_6 / a_0$ & $E_6$ (kHz) & $R_\mathrm{dip}^\mathrm{lim}/R_6$ & $F_\mathrm{0.7}$ (kV/cm) \\
\hline
KRb  & 0.57 & $5.7 \times 10^3$ & 1725 & $2.4 \times 10^3$ & $154$ & 1200.6 & 37  & 22.2 \\
RbCs & 1.2  & $4.7 \times 10^4$ & 15.1 & $1.2 \times 10^5$ & $469$ & 74.6   & 99  & 4.5  \\
NaK  & 2.7  & $6.6 \times 10^4$ & 26.5 & $5.1 \times 10^5$ & $491$ & 237.2  & 134 & 11.7 \\
KCs  & 2.0  & $9.5 \times 10^4$ & 4.6  & $4.5 \times 10^5$ & $613$ & 55.9   & 155 & 5.1  \\
NaRb & 3.3  & $1.7 \times 10^5$ & 2.3  & $1.5 \times 10^6$ & $739$ & 60.0   & 229 & 7.1  \\
CaF  & 3.1  & $7.8 \times 10^5$ & 19.9 & $2.3 \times 10^5$ & $395$ & 391.3  & 198 & 37.7 \\
\hline\hline
\end{tabular}
\end{table*}

\begin{figure}
\begin{center}
\includegraphics[width=\lfig,clip]{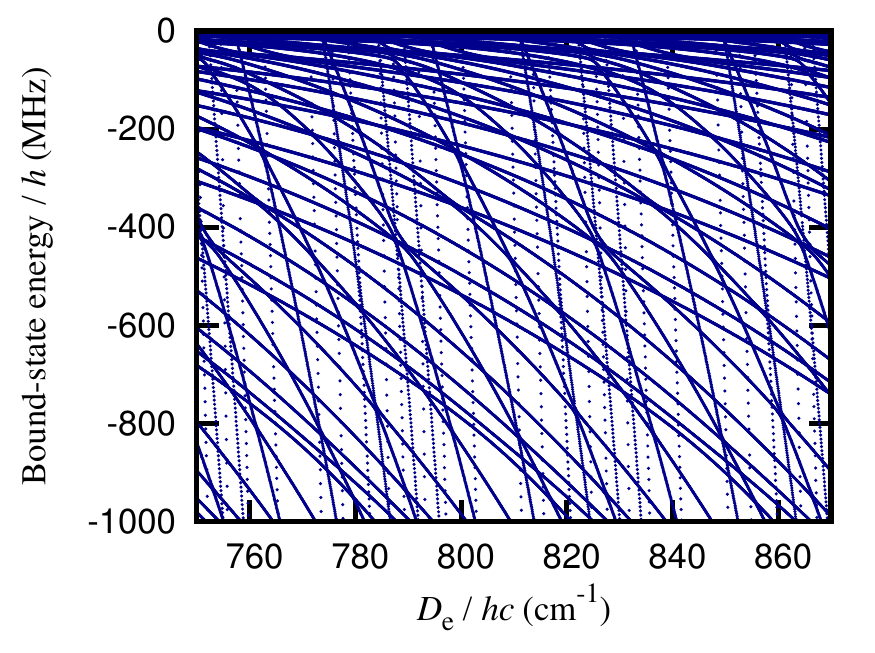}
\includegraphics[width=\lfig,clip]{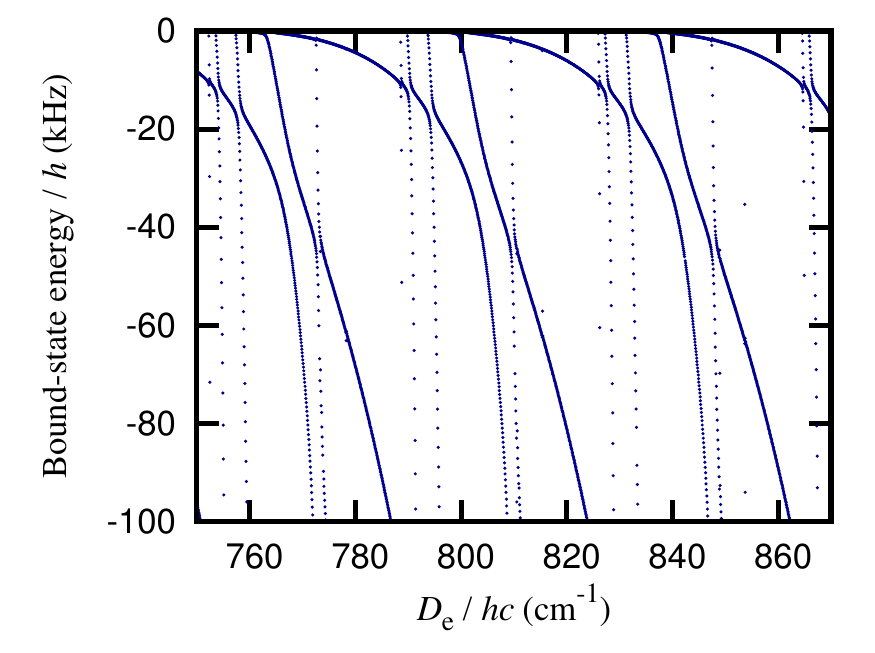}
\caption{ \label{fig:Dy10}
Dipole bound states with Lennard-Jones short-range interactions in the case
$R_\mathrm{dip} \gg R_6$. Top: On the same scale as Fig.~\ref{fig:Dy01}. Bottom: The vertical energy scale has been reduced by a
factor $10^4$.}
\end{center}
\end{figure}

Figure~\ref{fig:Dy10} shows the bound states as a function of the Lennard-Jones
well depth. The density of states is approximately proportional to $E^{-2/3}$.
The upper panel may be compared with the case of magnetic atoms in
Fig.~\ref{fig:Dy01}, where the density is lower and increases as about
$E^{-1/2}$ because the attractive part of the Lennard-Jones potential
dominates. The lower panel of Fig.~\ref{fig:Dy10} shows an expanded view of the
molecule-molecule bound states, with the energy range reduced by a factor of
$10^4$. In this range, there is a complicated level pattern with many avoided
crossings, similar to that seen with a hard-wall boundary condition in
Fig.~\ref{fig:BCT_bound_dip}. This contrasts with the case of magnetic atoms,
shown in Fig.~\ref{fig:Dy01}, where there are no significantly avoided
crossings close to threshold. Although the periodicity with $D_\mathrm{e}$ is
driven by the variation of the Lennard-Jones potential even in the molecular
case, it is clear that the dipole-dipole interaction is no longer simply
perturbative; instead, the Lennard-Jones effectively sets a short-range
boundary condition for the dipole-dipole interaction, but the latter is now
dominant and determines the structure of the states.

\begin{figure}
\begin{center}
\includegraphics[width=\lfig,clip]{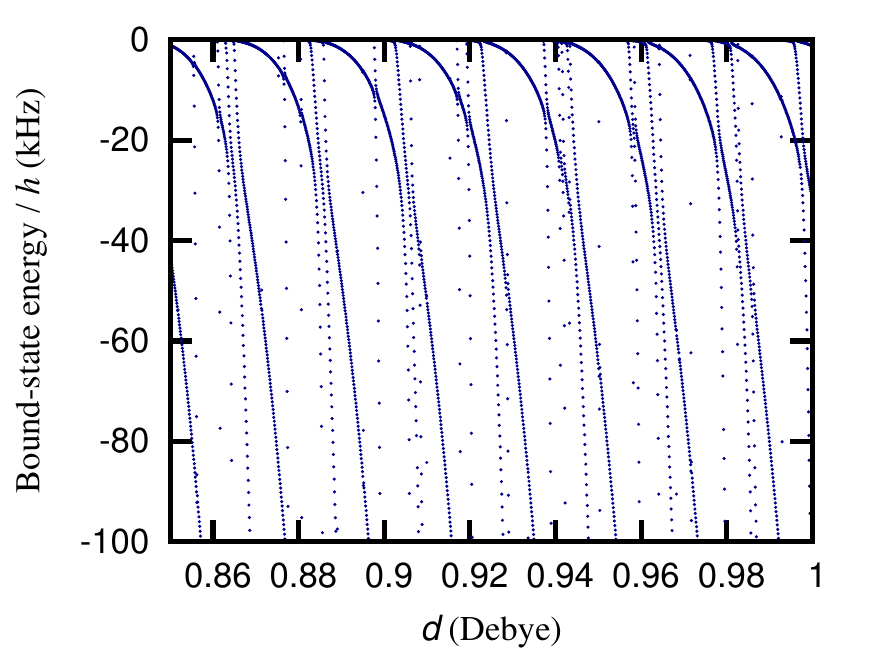}
\caption{ \label{fig:Dy10dip}
Dipole bound states with Lennard-Jones short-range interactions in the case
$R_\mathrm{dip} \gg R_6$, as a function of dipole moment for a fixed well depth.}
\end{center}
\end{figure}

Figure \ref{fig:Dy10dip} shows the states as a function of the space-fixed
dipole moment with the Lennard-Jones well depth fixed. The structure observed
is similar to that from the simpler calculations shown in
Fig.~\ref{fig:BCT_bound_dip}, which used a hard-wall short-range boundary
condition at a distance equal to the van der Waals length scale. However, the
states in Fig.~\ref{fig:Dy10dip} have an approximate period of about 0.02
Debye, which is considerably shorter than in Fig.~\ref{fig:BCT_bound_dip}. This
period corresponds to a hard wall around 7\,$a_0$, which is comparable to the
inner turning point of the Lennard-Jones potential. The lowest adiabat of the
dipole-dipole interaction at $R_6=154\,a_0$ is about $-500 E_6$, and at this
kinetic energy the transmission coefficient through the attractive part of the
Van der Waals potential is close to~1 \cite{gao:08}.

\begin{figure}
\begin{center}
\includegraphics[width=\lfig,clip]{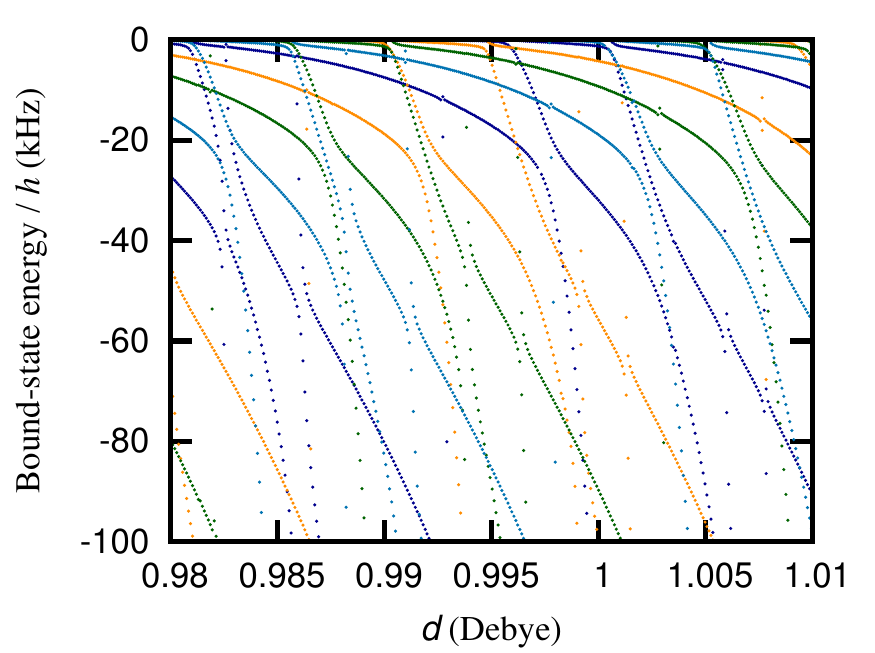}
\caption{ \label{fig:Dy10dip_var}
Dipole bound states with Lennard-Jones short-range interactions in the case
$R_\mathrm{dip} \gg R_6$, as a function of dipole moment for fixed well depths.
Different colors correspond to different well depths between roughly 800 and
830 cm$^{-1}$, which scans one cycle of scattering length.}
\end{center}
\end{figure}

Figure \ref{fig:Dy10dip_var} shows the states over a smaller range of dipole
moment for 4 different Lennard-Jones well depths over one period of the
Lennard-Jones scattering length. Although each of these show different details
in the states from higher adiabats, the overall dependence on the induced
dipole moment is similar for each of the different well depths and also similar
to that observed for hard-wall boundary conditions in Sec.~\ref{sec:hardwall}.
Since the pattern of states is independent of the short-range boundary
conditions, it should be present for real systems and should be observable
spectroscopically by measurements on trapped ultracold polar molecules.

\section{Conclusions \label{sec:conclusions}}

We have explored the bound states of a simple model of the dipole-dipole
interaction. The model assumes that both dipoles are oriented along a
space-fixed field direction. It has been used extensively in the description of
many-body physics with ultracold polar molecules and magnetic atoms, but its
two-body physics has been explored less fully. We have studied the bound states
of this model with both hard-wall boundary conditions and more realistic
Lennard-Jones short-range interactions.

In the simplest case, with hard-wall boundary conditions, we find a complicated
pattern of bound states that avoided-cross as a function of the boundary
condition or the space-fixed dipole moment. The pattern of states may be
understood using an adiabatic representation, which diagonalizes the
dipole-dipole interaction at each separation of the dipoles. States supported
by the lowest adiabatic potential curve approach threshold slowly as a function
of dipole moment, while those supported by excited adiabats approach more
steeply. The adiabatic approximation gives a good qualitative description of
the energy levels, except near avoided crossings between levels supported by
different adiabats.

The adiabats of a pure dipole-dipole potential are universal when expressed in
terms of the dipole length and dipole energy defined in Eq.\ \ref{eq:scales}.
For a pair of bosons, the lowest adiabat behaves as $-C_4 R^{-4}$ at very long
range. However, it deviates substantially from this form at distances around
the dipole length. Because of this, the $R^{-4}$ form accurately describes
states only if they are bound by much less than the dipole energy. The dipole
energy decreases very fast as the dipole itself increases, and for pairs of
polar molecules is typically less than 1~kHz $\times h$. This is so shallow
that there are usually no states in the region characterized by the asymptotic
$R^{-4}$ form.

For fermions, with odd $L$, states with $M_L=0$ are almost identical to the
boson states when bound by more than the dipole energy. This may also be
understood through the adiabatic representation, as the states can localize
with dipoles head-to-tail in both cases. There are also fermion states with
$|M_L|>0$, which are confined by a higher centrifugal barrier.

We have also considered two cases with a Lennard-Jones short-range potential in
place of the hard-wall boundary condition. When the dipolar length scale is
comparable to the van der Waals length scale, as is the case for magnetic atoms
with weak dipolar interactions, the bound states are dominated by the
Lennard-Jones potential, and the dipole-dipole interaction acts perturbatively.
The dipole-dipole coupling is suppressed by the structure of van der Waals
bound states, which groups together states with $\Delta L \ge 4$ that are not
coupled directly by dipole-dipole interactions.

When the dipolar length scale is much larger than the van der Waals length
scale, as is achievable for polar molecules with large dipole moments, there is
a denser set of dipole-dipole bound states close to threshold. These states can
be tuned across threshold by varying the dipole moment with an applied electric
field and exhibit complex patterns of avoided crossings below threshold.
Spectroscopic measurements of these bound states, and observation of resonances
where they cross threshold, have great potential to help understand dipolar
interactions and illuminate their role in few-body and many-body quantum
systems.

\bibliography{bibfile,../all}
\end{document}